\documentclass[3p,11pt]{elsarticle}

\usepackage{amsmath}
\usepackage{graphics}
\usepackage{color}
\usepackage{array}
\usepackage{booktabs}
\usepackage{rotating}
\usepackage{hyperref}

\newcolumntype{P}[2]{%
  >{\begin{turn}{#1}\begin{minipage}{#2}\small\raggedright\hspace{0pt}}l%
  <{\end{minipage}\end{turn}}%
}

\journal{Physica A}

\begin{document}

\begin{frontmatter}

\title{Characterization of river flow fluctuations via horizontal visibility graphs}
\author[a,b]{A.~C.~Braga}
\author[c]{L.~G.~A.~Alves}
\author[c]{L.~S.~Costa}
\author[b]{A.~A.~Ribeiro}
\author[d]{M.~M.~A.~de~Jesus}
\author[e]{A.~A.~Tateishi}
\author[c,d]{H.~V.~Ribeiro}
\ead{hvr@dfi.uem.br} 

\address[a]{Departamento de Matem\'atica, Universidade Tecnol\'ogica Federal do Paran\'a, Apucarana, PR 86812-460, Brazil}
\address[b]{Departamento de Matem\'atica, Universidade Federal do Paran\'a, Curitiba, PR 81531-980, Brazil}
\address[c]{Departamento de F\'isica, Universidade Estadual de Maring\'a - Maring\'a, PR 87020-900, Brazil}
\address[d]{Departamento de F\'isica, Universidade Tecnol\'ogica Federal do Paran\'a - Apucarana, PR 86812-460, Brazil}
\address[e]{Departamento de F\'isica, Universidade Tecnol\'ogica Federal do Paran\'a - Pato Branco, PR 85503-390, Brazil}

\begin{abstract}
We report on a large-scale characterization of river discharges by employing the network framework of the horizontal visibility graph. By mapping daily time series from 141 different stations of 53 Brazilian rivers into complex networks, we present an useful approach for investigating the dynamics of river flows. We verified that the degree distributions of these networks were well described by exponential functions, where the characteristic exponents are almost always larger than the value obtained for random time series. The faster-than-random decay of the degree distributions is an another evidence that the fluctuation dynamics underlying the river discharges has a long-range correlated nature. We further investigated the evolution of the river discharges by tracking the values of the characteristic exponents (of the degree distribution) and the global clustering coefficients of the networks over the years. We show that the river discharges in several stations have evolved to become more or less correlated (and displaying more or less complex internal network structures) over the years, a behavior that could be related to changes in the climate system and other man-made phenomena.
\end{abstract}

\end{frontmatter}
\clearpage
\section*{Introduction}

The study of earth-related systems has become even more important with the growing concerns about environmental changes and the awareness of sustainable development. As a paradigm of complex systems, this research topic relies on multidisciplinary efforts and has also been addressed by physicists via methods of statistical physics. Earthquakes~\cite{Mendes,Ribeiro2}, geomagnetic activities~\cite{Turner}, climate~\cite{Boettle} and weather-related systems~\cite{Rybski,Ribeiro}  are just a few examples of systems that researchers have tackled in these pages. In particular, as pointed out by Dove and Kammen~\cite{Dove}, one of the foremost global environmental challenges is the climate change. In a broader sense, the complexity of climate systems is related to the complex interactions between atmosphere, biosphere, cryosphere, lithosphere and  hydrosphere. The latter one is the part of the climate system that comprises oceans, lakes and rivers, that is, the liquid water at the Earth's surface and underground~\cite{WMO} and it is well known the extremely important role of water in global environmental change~\cite{Oliver,Machiwal}.

In this context, important systems are the rivers and their discharges, which have a large impact on human activities, and that may also suffer huge influence from these activities. A river flow results from complicated interactions between the weather-related systems (such as rainfall, temperature and evaporation), the landscape (such as basin area and land relief) and human activity (such as pollution and power generation). These many features make river flow rates (river discharges) a complex process that has attracted the attention of scholars over the last six decades. For instance, the seminal work of Hurst about the long-range dependence of runoff records from several rivers~\cite{Hurst} has fostered several discussions on the fractal/multifractal and scaling properties of the temporal evolution of river flows~\cite{Hosking,Hipel,Montanari2,Koutsoyiannis,Dahlstedt,Wang,Dolgonosov,Movaheda,Zhang,Zhang2,Montanari3,Domenico,Yu,Montanari}. The correlations between river flows and other  systems have been also studied, for instance, climate systems such as rain fall~\cite{Tessier, Kantelhardt2,Bogachev} and sunspots~\cite{Hajian}; and economic systems such as the growth of companies~\cite{Janosi,Bramwell}. Moreover, chaos theory~\cite{Porporato,Bordignon}, stochastic models~\cite{Livina} and  permutation entropy~\cite{Mihailovic,Hauhs,Zunino2,Lange,Serinaldi} are examples of approaches used to probe the complexity of runoff time series. 

Despite the considerable attention towards the investigation of river flows, several works are still based on small datasets and a large-scale characterization of time series related to river flows is rarely reported. Furthermore, previous efforts have been mainly focused on well established/traditional methods of times series analysis (such as fractal/multifractal analysis) and, for instance, the interesting advances in mapping time series into networks have just recently attracted the attention of researchers working on this topic~\cite{Jha,Scarsoglio,Sivakumar1,Sivakumar2,Sivakumar3}. Here, we further fill this gap by studying the flows in 141 different measuring stations that cover 53 Brazilian rivers via daily time series obtained from the period of 1931 to 2012. Specifically, we have employed the network framework of the horizontal visibility graph~\cite{Luque,Lacasa,Lacasa2} for mapping these river flow series into networks. By tracking the evolution of topological properties of these networks, the horizontal visibility approach reveals that the flow in several stations are becoming more or less correlated (and displaying more or less complex internal network structures) over the years, a behavior that could be related to changes in the climate system and other man-made phenomena.


This work is organized as follows. We first describe our database and review some properties of the horizontal visibility approach. We next employ the horizontal visibility graph to our time series. Then, we characterize these networks by investigating the degree distribution and the clustering coefficient as well as evolutive features of these two measures. We also find that these measures display a kind of coupling. Finally, we present a summary of our findings and some concluding remarks. 

\section*{Data presentation and analysis}
\begin{figure*}[!ht]
\centering
\includegraphics[scale=0.45]{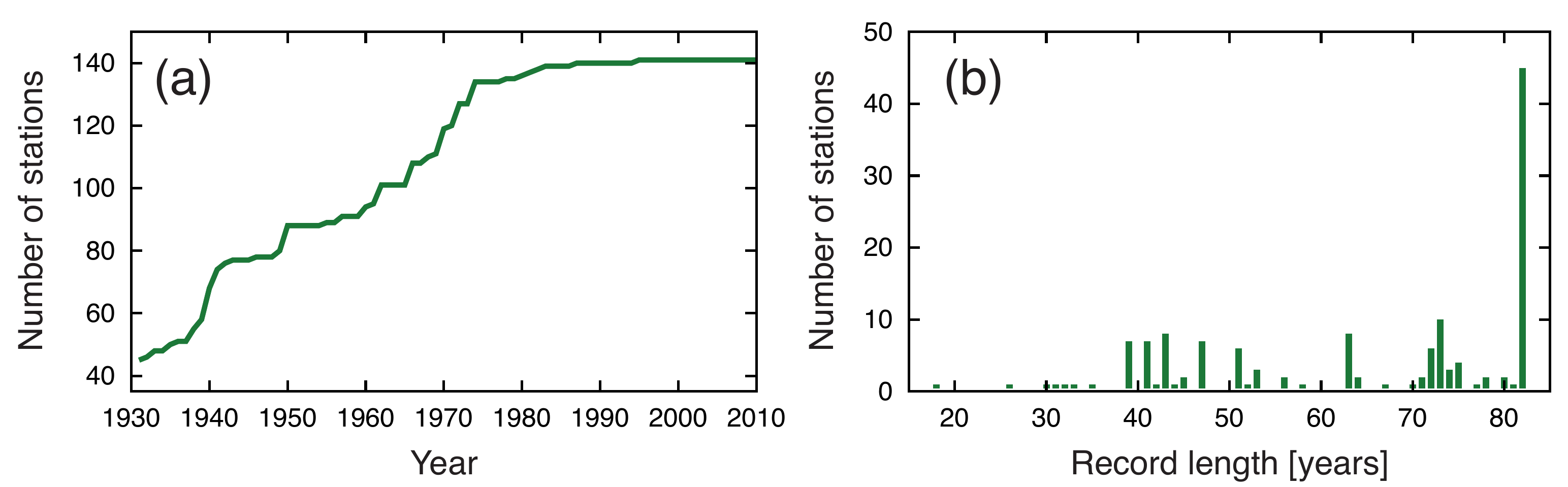}
\caption{(color online) Schematic description of the dataset. (a) Number of stations over the years. The dataset starts with 45 stations in 1931 and 
since 1995 there are 141 stations covering 53 Brazilian rivers. (b) Histogram of the record length (in years) for all stations.
}
\label{fig:0}
\end{figure*}

The data we have accessed consist of time series of the natural river flow rates (river discharges) with daily resolution measured in 141 different stations. These time series cover 53 Brazilian rivers, span the period from 1931 to 2012 [as described in Fig.~\ref{fig:0}(a) and (b)] and are made freely available by the Operador Nacional do Sistema El\'etrico --- ONS --- (a federal institution that controls the power system in Brazil)~\cite{ONS}. Let's denote the flow rates by $F_t(i)$, where $i=1,2,\dots,365$ is a discrete time variable indexing the days of the year and $t$ stands for the year associated with the flow; thus, $F_{1986}(10)$ represents the flow rate in Jan. 10 1986 in a given station. For the matter of convenience, we have removed the datapoint associated to Feb. 29 from all time series of leap years. 

\begin{figure*}[!ht]
\centering
\includegraphics[scale=0.35]{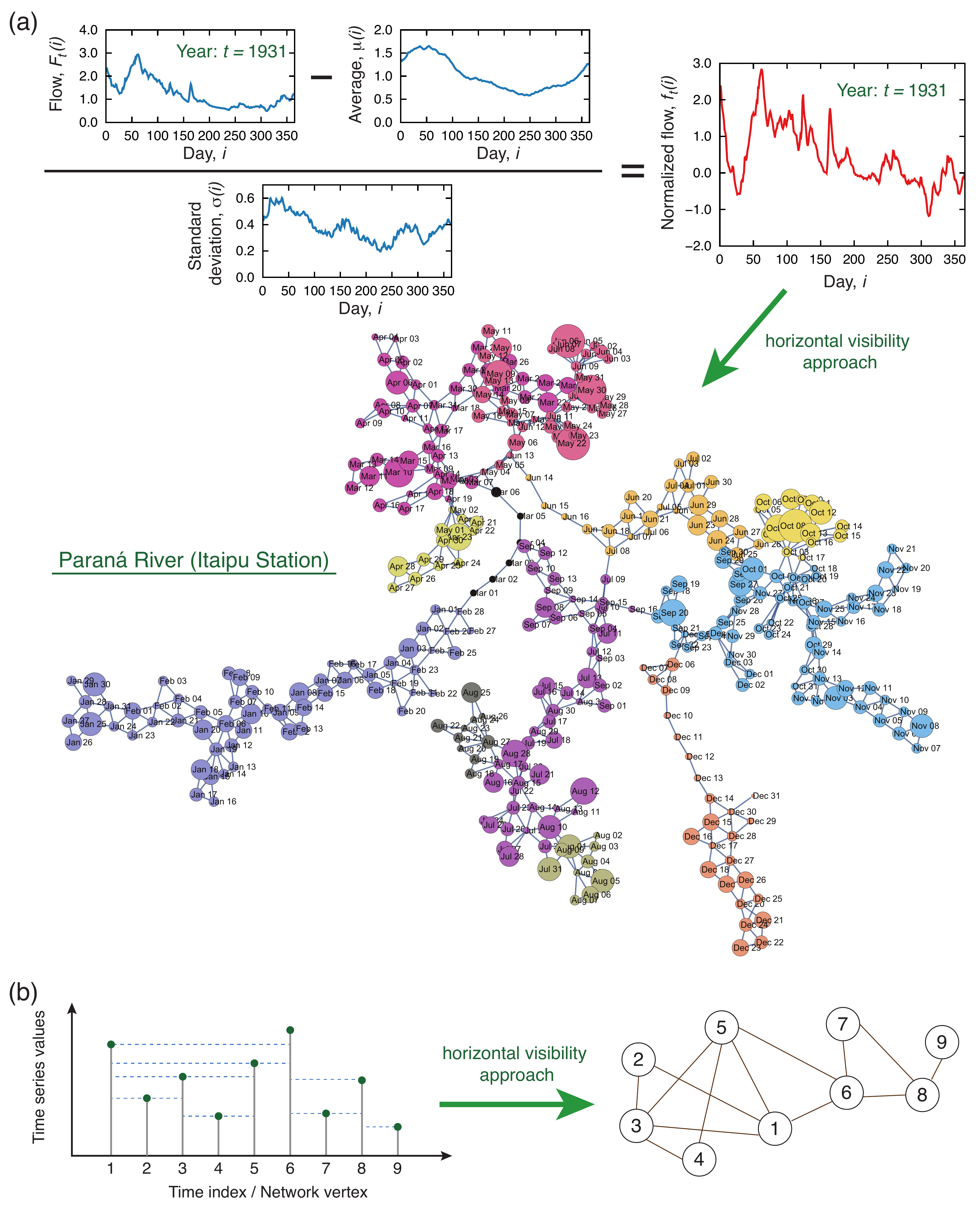}
\caption{(color online) Schematic illustration of the network construction from the river discharges. (a) The upper panel illustrates the definition of the normalized flow $f_t(i)$ over the days $(i=1,2,\dots,365)$ for a given year and station. The procedure consists in subtracting the daily average flow $\mu(i)$ from the flow $F_t(i)$ in a given year and dividing the result by the daily standard deviation flow $\sigma(i)$ (all in units of $10^4~\times$~m$^{3}/$s ). Once $f_t(i)$ is defined, we employ the horizontal visibility approach for building up the network, where each node represents a day of the year (shown in the network). This particular network has been built with data from the Paran\'a River measured in the Itaipu Station in the year $t=1931$. The size of nodes are proportional to their degrees and the colors are just illustrative. (b) Illustrative application of the horizontal visibility graph approach. The left panel shows a simple time series represented by vertical bars. Here,  the horizontal dashed lines indicate the network connections established according the geometrical criterion of Eq.~\ref{eq:hvg}. The right panel shows the network emerging from this time series.}
\label{fig:1}
\end{figure*}

We have focused our analysis on a normalized version of the flow rates defined as
\begin{equation}
f_t(i) = \frac{F_t(i)-\mu(i)}{\sigma(i)}\,,
\end{equation}
where
\begin{equation}
\mu(i) = \frac{1}{n}\sum_{t=1}^{n} F_t(i)~~\text{and}~~\sigma^2(i) = \frac{1}{n-1}\sum_{t=1}^{n} [F_t(i) - \mu(i)]^2
\end{equation}
are respectively the average and the variance flow profile along the days of the year for a given station ($n$ is the number of years available for that station). The upper panel of Fig.~\ref{fig:1}(a) illustrates the definition of the normalized flow rates $f_t(i)$ by showing a concrete example of its construction. By doing this procedure, we ensure that (at least) the main seasonal trend is removed from the original flow rates.

Once having the normalized flow $f_t(i)$, we have applied the horizontal visibility graph approach for mapping each year of the time series in a corresponding complex network. This procedure was proposed by Luque et al.~\cite{Luque} as a more restrictive version of the visibility graph approach~\cite{Lacasa} with the advantage of providing analytical expressions for fully random time series~\cite{Luque,Lacasa2}. The visibility approach has been applied to several contexts~\cite{Lacasa3,Yang,Elsner,Ahmadlou,Murks,Telesca,Gao,Jiang,Zhuang,Telesca2,Zou,Zhangvg}, enabling researchers from time series analysis to employ tools from network science. Several properties of the horizontal visibility graph can be found in Refs.~\cite{Luque,Lacasa2}; here, we briefly present its algorithm and some results related to the degree distribution. 

The horizontal visibility algorithm is a map that assigns each datapoint of a time series to a node/vertex in a complex network. Two nodes, \textcircled{\small{$i$}} and \textcircled{\small{$j$}}, will be connected whenever one can draw a horizontal line in the time series space that join the datapoints $f_t(i)$ and $f_t(j)$ without intersecting any intermediate datapoint height, that is,
\begin{equation}\label{eq:hvg}
\textcircled{\small{$i$}} \leftrightarrow \textcircled{\small{$j$}}~~\text{when}~~ [f_t(i),f_t(j)] > f_t(l)~\forall\,l\, | (i<l<j)\,.
\end{equation}
Notice that, in our case, the horizontal visibility algorithm will produce one network per year of the time series of a given station. The bottom panel of Fig.~\ref{fig:1}(a) shows an example of a network constructed from the previous procedure by considering data from the Paran\'a River collected at the Itaipu Station in the year $t=1931$. Also, Fig.~\ref{fig:1}(b) illustrate the horizontal visibility approach applied to a simple time series. Thus, our approach enables the investigation of possible evolutive features of the river flow rates by tracking the evolution of network measures. 

We first evaluated the degree distribution of the networks $P(k)$, where $k$ stands for the node degree or its number of connections (notice that the networks are undirected). From the works of Luque~\cite{Luque} and Lacasa~\cite{Lacasa2}, we know that the degree distribution takes the form
\begin{equation}
P(k)\sim \exp(-\lambda\, k)~~\text{with}~~\lambda=\lambda_{\text{rand}}=\ln(3/2)
\end{equation}
for fully random time series, regardless the probability distribution underlying the values of the time series. Furthermore, time series arising from more complex dynamics are also usually described by asymptotic exponential distributions. However, the value of $\lambda$ is usually different from $\lambda_{\text{rand}}$. Actually, $\lambda<\lambda_{\text{rand}}$ is associated with chaotic processes (the smaller the $\lambda$, the smaller the system dimensionality) and $\lambda>\lambda_{\text{rand}}$ is associated with stochastic processes (the larger the $\lambda$, the longer the system correlations)~\cite{Lacasa2}. 

\begin{figure*}[!ht]
\centering
\includegraphics[scale=0.3]{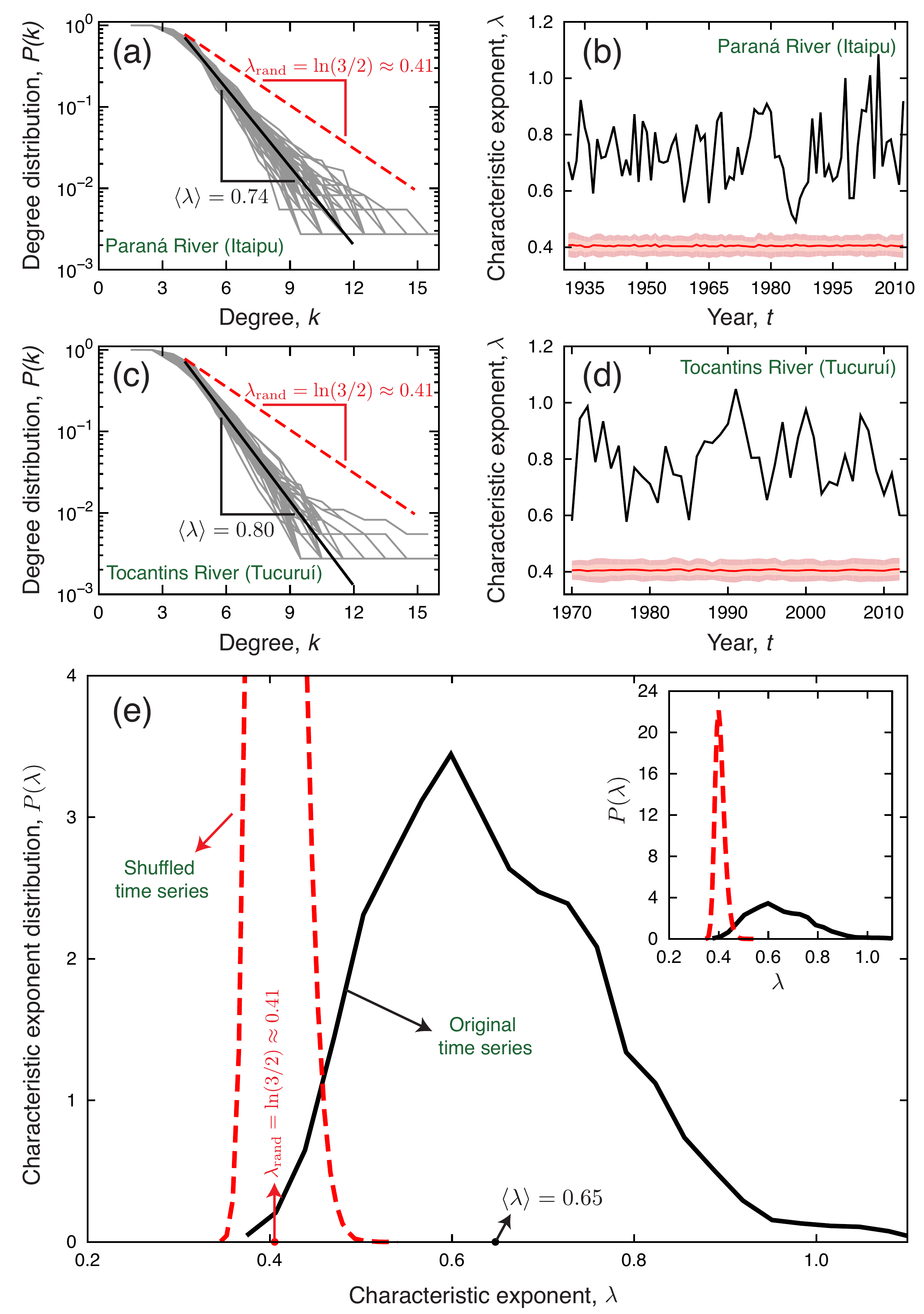}
\caption{(color online) Degree distributions and the correlated nature of the normalized river discharges. (a) Cumulative distributions of the vertex degree $P(k)$ (log--lin scale) for each network built from time series available for the Paran\'a River (Itaipu Station). The gray curves are the distributions for each year and the solid black line is an exponential fit to the window average of these distributions, where the average characteristic exponent is $\langle\lambda \rangle = 0.74\pm 0.01$. The red dashed line illustrates the exponential decay expected for a random time series, that is, $P(k)\sim \exp(-k/\lambda_{\text{rand}})$, with $\lambda_{\text{rand}}=\ln(3/2)\approx0.41$. (b) Evolution of the characteristic exponent $\lambda$ obtained for each one of the previous distributions. The black line shows the empirical values of $\lambda$ for each year. The red line shows the values  $\lambda$ after randomizing the time series (averaged over 100 realizations) and the light (dark) shaded areas stands for their 95\% (99\%) bootstrap confidence intervals. Panel (c) and (d) show the same quantities for the Tocantins River (Tucuru\'i Station). (e) Distribution $P(\lambda)$ for the values of $\lambda$ obtained from the original times series (black line) and for the randomly shuffled ones (red line). Notice that the values of $\lambda$ are concentrated around $\lambda_{\text{rand}}$ for the randomized series; whereas the values of $\lambda$ of the original times series are almost always larger than $\lambda_{\text{rand}}$ ($99.7\%$ of the series), having an average value (evaluated over all rivers and years) equal to $0.65$. The inset shows these distributions for a larger plot range.
}
\label{fig:2}
\end{figure*}

In our case, Fig.~\ref{fig:2}(a) shows the cumulative degree distributions for all the 82 networks (one for each year) constructed with data from the Paran\'a River at the Itaipu Station. We observe that all distributions are asymptotically well described by exponential decays, which on log--lin scale are represented by straight lines whose slopes match the values of $\lambda$. When comparing the asymptotic behavior of these distributions to the one expected for a random time series [$P(k)\sim \exp(-\lambda_{\text{rand}}\, k)$], we note that the empirical distributions display a faster decay. In order to numerically estimate the empirical value of $\lambda$ for each distribution, we have fitted a linear model to each cumulative distribution (on log--lin scale) after removing the initial non-exponential behavior ($k>4$). We have further obtained the average value of $\lambda$ after shuffling the times series over 100 realizations. Figure~\ref{fig:2}(b) shows the values of $\lambda$ for each year of the original series as well as the average (and confidence intervals) of $\lambda$ for the shuffled versions of these series. We observe that the empirical values of $\lambda$ (for this particular station) are always larger than $\lambda_{\text{rand}}$ and that they are outside the confidence bounds related to random versions of these series. Figures~\ref{fig:2}(c) and~(d) provide another representative example (for the Tocantins River at the Tucuru\'i station) of this analysis, where again we have found values of $\lambda$ larger than $\lambda_{\text{rand}}$.

In order to fully characterize the degree distributions of our entire dataset, we have proceeded as in the two previous-discussed examples for estimating the values of $\lambda$ for every time series. After, we calculate the probability distribution for the values of $\lambda$ obtained from the original time series as well as for one shuffled version of each time series. Figure~\ref{fig:2}(e) shows both distributions, where it is evident that the values of $\lambda$ for the original series are larger than $\lambda_{\text{rand}}$ for practically all time series in our dataset, only $0.3\%$ of the time series display slightly smaller values (but close to $\lambda_{\text{rand}}$). It is worth noting that the average value of $\lambda$ is 0.65, a value that can be related to long-range correlations in the river flows. In particular, if consider that our time series are described by a power-law correlation function $R(\tau)\sim\langle f_t(i+\tau) f_t(i)\rangle \sim\tau^{-\gamma}$, an average $\langle\lambda \rangle=0.65$ corresponds to $\gamma\approx0.5$ (see figure 3 of Ref.~\cite{Lacasa2}) and consequently to a Hurst exponent around 0.75, a result that is compatible with the existence of long-range persistent correlations in the river flows.


\begin{figure*}[!ht]
\centering
\includegraphics[scale=0.54]{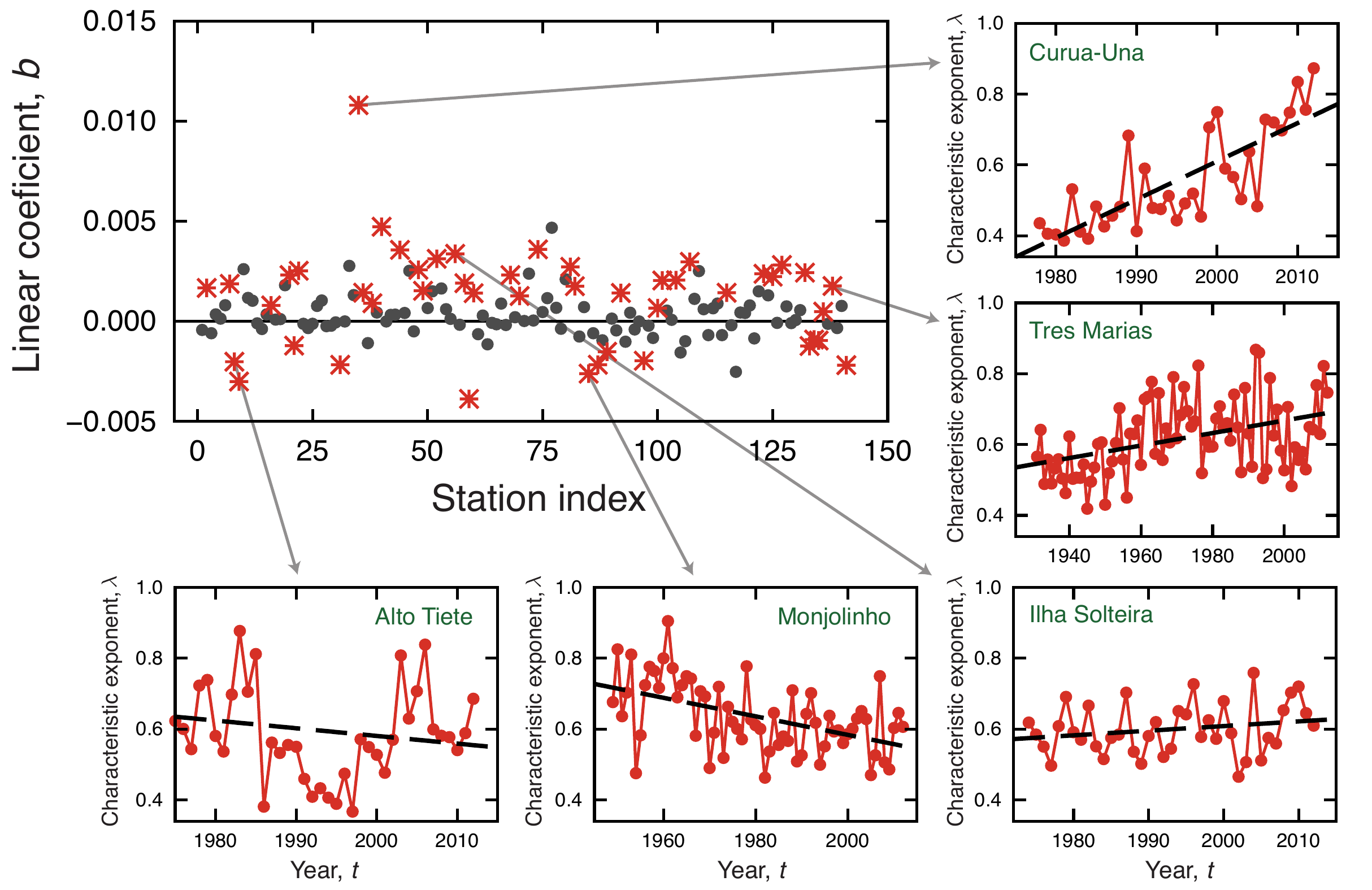}
\caption{(color online) Evolutive trends in the degree distribution. The main panel shows the values of the linear coefficients $b$ obtained by least squares fitting the model $\lambda = a + b\,t$ to the relationships between the characteristic exponent $\lambda$ and $t$ (the year associated with the time series) for all rivers. The gray circles show the values of $b$ that are not statistically significant (that is, rivers in which the relationship between $\lambda$ and $t$ are well approximated by a constant function) and the red asterisks show the significant ones. The panels indicated by the arrows provide representative cases for the relationship $\lambda$ versus $t$ (red circles), where the dashed lines represent the adjusted linear models.
}
\label{fig:3}
\end{figure*}

Another question that our data pose is regarding possible evolutive features in the river flow dynamics. To address this question, we have investigated whether the values of $\lambda$ of a given station display a time dependence over the years $t$. Specifically, we have tested the hypothesis of the relationship $\lambda$ versus $t$ showing a linear trend by fitting the linear model (via the ordinary least squares method)
\begin{equation}
\lambda = a + b\, t\,,
\end{equation}
where $a$ and $b$ are the linear coefficients. Thus, the value of $b$ and its statistical significance provide clues of whether $\lambda$ is changing over the years $t$ or not. Figure~\ref{fig:3} shows the value of $b$ for each station. In order to check the statistical significance of $b$, we have employed the $t$-test for examining the null hypothesis of $b$ being different from zero via the two-tail $p$-value. For a confidence level of 95\% (that is, $p$-value smaller than 0.05), we find that the null hypothesis cannot be rejected in 46 among 141 measuring stations (Fig.~\ref{fig:3}). Also, among the stations where $b$ is statistically significant, we find that the majority ($72$\%) displays an increasing trend. This result thus indicates that some stations are presenting a more correlated dynamics over the years. It is important to mention that the $t$-test assumes that the residual distributions are normal and it can be considered not ideal in this context~\cite{Koutsoyiannis2}. However, we have further applied the bootstrap regression approach~\cite{Efron} and the results obtained are quite similar, that is, the linear trends considered significant by the $t$-test, were also considered statistically significant by the bootstrap regression.

\begin{figure*}[!ht]
\centering
\includegraphics[scale=0.3]{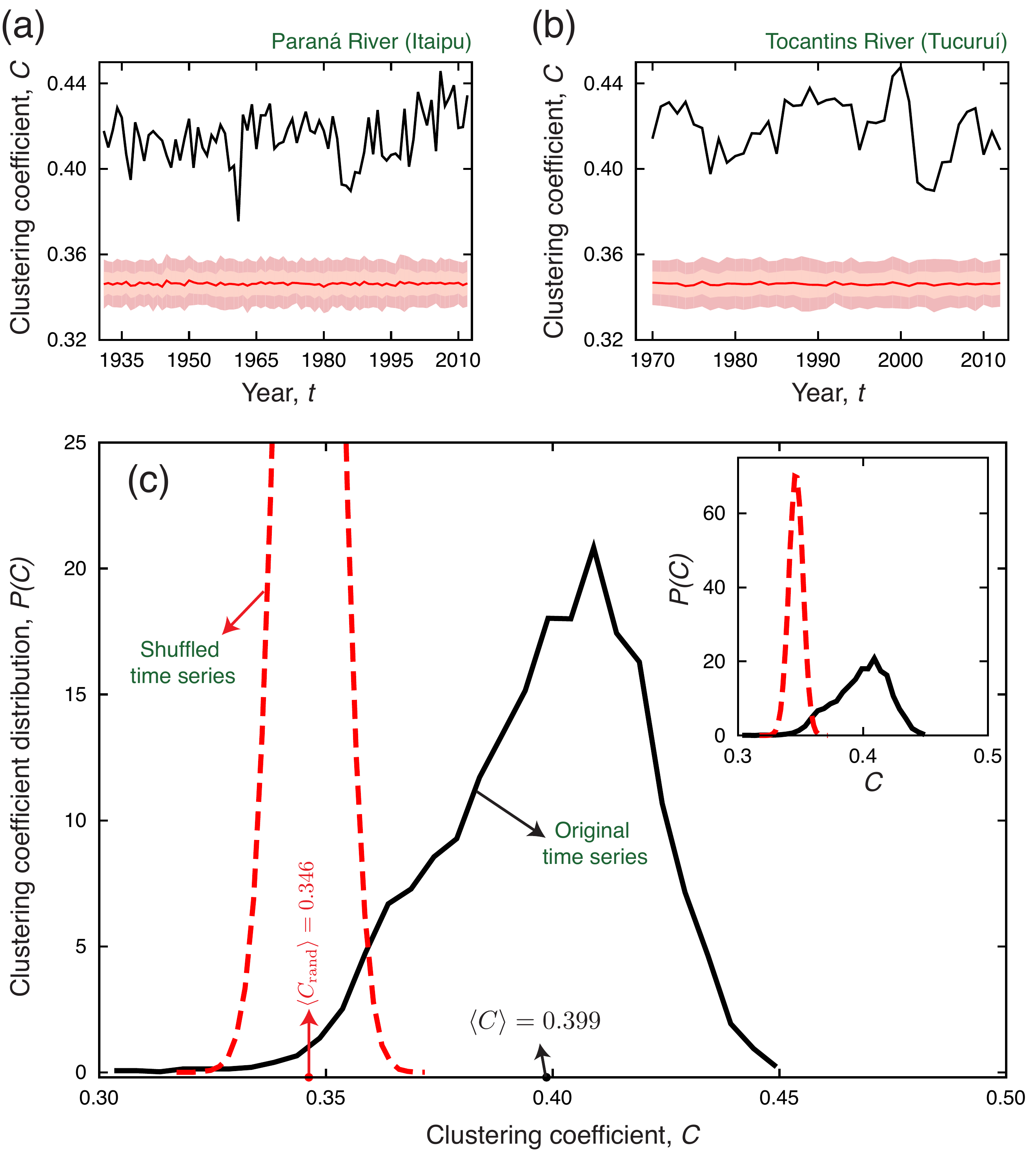}
\caption{(color online) Global clustering coefficients of the networks. Evolution of the clustering coefficient $C$ for each network built from the time series available for (a) the Paran\'a River (Itaipu Station) and (b) the Tocantins River (Tucuru\'i Station). The black lines show the values of $C$ for each year $t$. The red lines show the values of $C$ after randomizing the time series (averaged over 100 realizations) and the light (dark) shaded areas stands for their 95\% (99\%) bootstrap confidence intervals. (c) Clustering coefficient distribution $P(C)$ for the values of $C$ obtained from the original times series (black line) and the randomly shuffled ones (red line). Notice that the values of $\lambda$ are concentrated around $\langle C_{\text{rand}}\rangle=0.346$ for the randomized series; whereas the values of $C$ from the original times series are almost always larger than $\langle C_{\text{rand}}\rangle$ ($99.1\%$ of the series), having an average value (evaluated over all rivers and years) of $0.399$. The inset shows these distributions for a larger plot range. 
}
\label{fig:4}
\end{figure*}

In addition to the degree distribution, another common-studied network property is the clustering coefficient~\cite{Watts,Newman,Newman2}. This quantity measures the likelihood of nodes to create tight-knit groups with a relatively high density of ties. Specifically, we have employed the global clustering coefficient $C$, which is the number of closed paths of length two in the network over the number of all paths of length two. Similarly to the analysis of the degree distribution, we have compared the values of $C$ obtained from the networks emerging from the original time series with the networks obtained from shuffled versions of these series (average over 100 realizations). Figures~\ref{fig:4}(a) and~\ref{fig:4}(b) show representative cases for the values of $C$ evaluated from two measuring stations. For both stations, we note that the values of $C$ evaluated for the original time series (black lines) are considerably larger than the values obtained from the randomized series (red lines). Similar behaviors are observed for almost all time series. In fact, we note from Fig.~\ref{fig:4}(c) that the values of $C$ from the original series are distributed around $\langle C \rangle = 0.399$, whereas the values of $C$ related to the shuffled series present a more acute distribution around $\langle C_{\text{rand}} \rangle = 0.346$. This result thus indicates that the underlying fluctuations of the river flows produce networks with more complex internal structures, which are visually observed in Fig.~\ref{fig:1} by noticing the formation of cliques among closer days.

\begin{figure*}[!ht]
\centering
\includegraphics[scale=0.54]{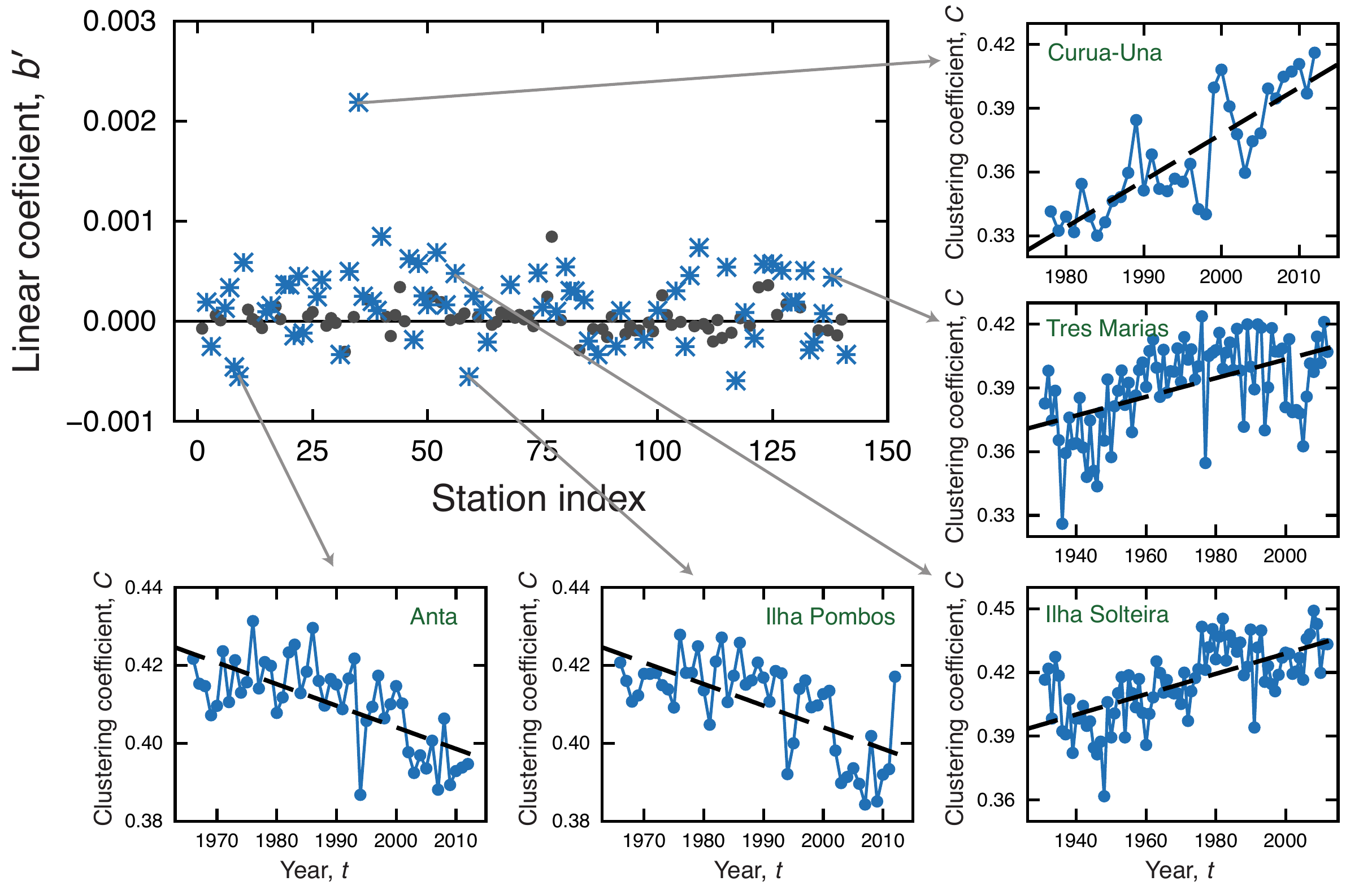}
\caption{(color online) Evolutive trends in the clustering coefficient. The main panel shows the values of the linear coefficients $b^\prime$ obtained by least squares fitting the model $C = a^\prime + b^\prime\,t$ to the relationships between the clustering coefficient $C$ and $t$ (the year associated with the time series) for all rivers. The gray circles show the values of $b^\prime$ that are not statistically significant (that is, rivers for which the relationship between $C$ and $t$ are well approximated by a constant function) and the blue asterisks show the significant ones. The panels indicated by the arrows provide representative cases for the relationship $C$ versus $t$ (blue circles), where the dashed lines represent the adjusted linear models. Notice that most of the stations displaying significative evolutions for the $C$ values also present evolutive trends in the $\lambda$ values (see Fig.~\ref{fig:3}). 
}
\label{fig:5}
\end{figure*}

We have also investigated the evolutive features in the values of $C$ over the years $t$ for all stations. As we did previously for the degree distribution, we have verified the hypothesis of $C$ versus $t$ displaying a linear trend by least squares fitting the model
\begin{equation}
C = a^\prime+b^\prime\, t\,,
\end{equation}
where $a^\prime$ and $b^\prime$ are now the model parameters. Again, the value of $b^\prime$ and its significance provide clues of whether $C$ is changing over the years $t$ or not. Figure~\ref{fig:5} shows the values of $b^\prime$ for each station, where the asterisk markers indicate the values that are statistically significant (confidence level of 95\%). Among the 141 measuring stations, we find out that 68 stations exhibit a statistically significant linear trend and that the majority of these stations ($72$\%) present an increasing trend in $C$ (similar results are obtained with the bootstrap regression approach). We further show in Fig.~\ref{fig:5} some representative cases of these evolutive trends. 

\begin{figure*}[!ht]
\centering
\includegraphics[scale=0.4]{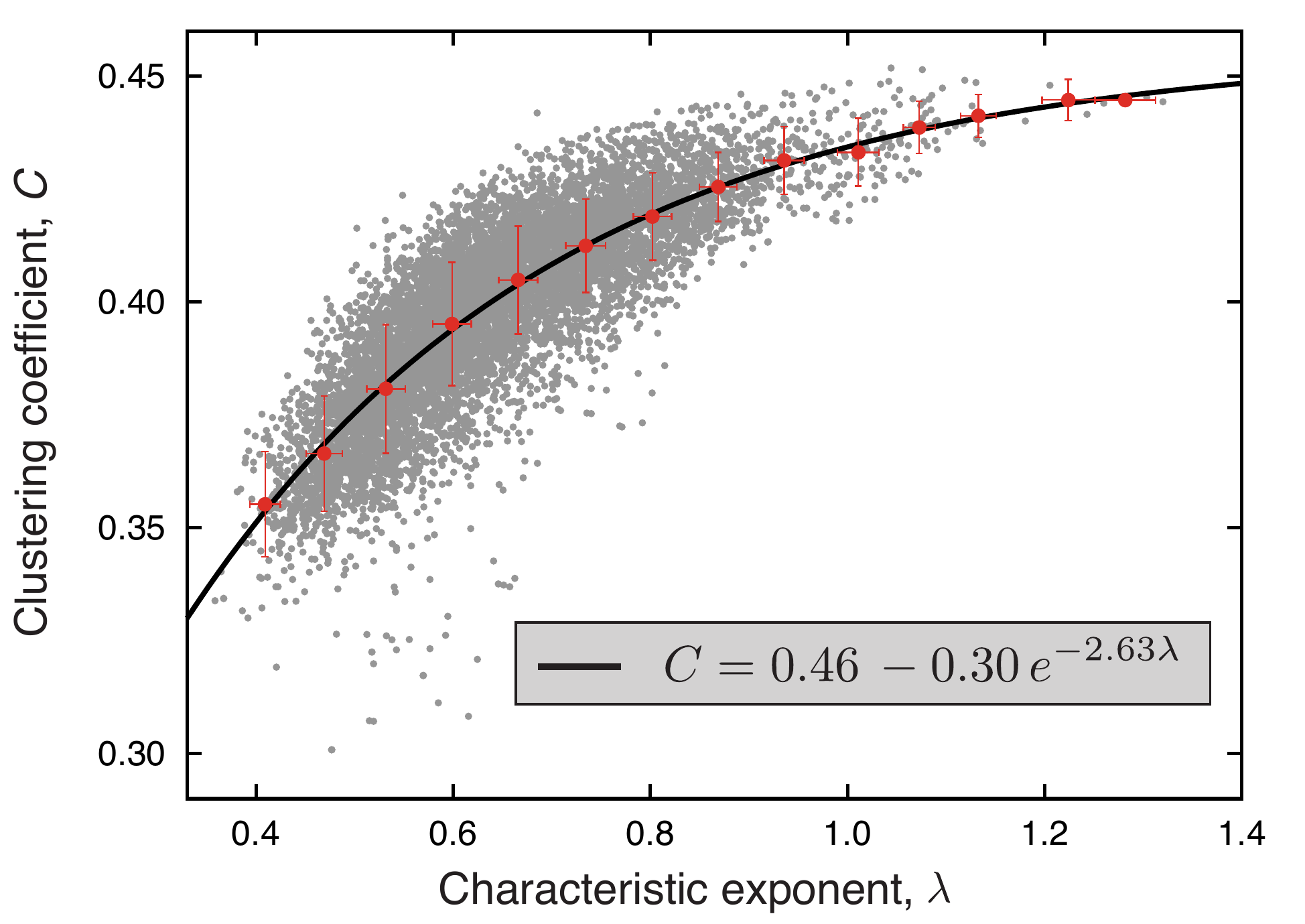}
\caption{The relationship between the clustering coefficient and the characteristic exponent. The gray dots are the values of the clustering coefficient $C$ versus the characteristic exponent $\lambda$ for each time series in our dataset. The red dots are window average values of the previous relationship, error bars stand for one standard deviation, and the solid line is an exponential fit to average values (parameters are shown in the plots).} 
\label{fig:6}
\end{figure*}

Another intriguing aspect of the Fig.~\ref{fig:5} is that most of the stations displaying evolutive features in $C$ have also presented a similar dynamics in $\lambda$. This behavior suggests that the values of $C$ and $\lambda$ may be somehow coupled. In order to check this possibility, we have plotted the values of $C$ against the values of $\lambda$ evaluated for each station and year of the time series. Figure~\ref{fig:6} shows this relationship where (despite the scatter) we note that the increasing of $\lambda$ is (on average) followed by an increasing of $C$. In order to overcome the noise and to focus on the main tendency of these data, we have calculated the window average values of this relationship. Our results suggest that an exponential function describes quite well the average relationship, which confirms a coupling (on average) between the values of $C$ and $\lambda$. 

\section*{Summary and Conclusions}
We have studied the river discharges from a relatively large dataset composed of almost one hundred and fifty stations covering more than fifty Brazilian rivers in time periods of more than eighty years. In our investigations, we have proposed to employ the network framework of the horizontal visibility method for characterizing river flow fluctuations. Our approach has pointed some intriguing evolutive features of the river discharges. We have shown that river discharges in certain measuring stations are becoming more or less correlated as well as displaying more or less complex internal network structures. Despite the difficult in testing this hypothesis from our data, we believe that these evolutive features may be related to changes in the climate system, in particular regarding the rainfall system. Other man-made phenomena such as the large-scale use of water in agricultural activities may also have contributed to these evolutive features. Perhaps, other investigations at more local level could help to elucidate the mechanisms underlying the evolutive aspects of the river discharges presented here. Finally, we believe that our work sheds new light on river flow dynamics and open the possibility for other direct investigations based on the horizontal visibility approach. 

\section*{Acknowledgements}
This work has been supported by the CNPq, CAPES and Funda\c{c}\~ao Arauc\'aria (Brazilian agencies). HVR thanks the financial support of the CNPq under Grant No.~440650/2014-3. AAT thanks the financial support of the CNPq under Grant No.~462067/2014-9. The authors are also grateful to Dr. Francesco Serinaldi for helpful discussions during the review process of this manuscript.


\begin{thebibliography}{99}

\bibitem{Mendes} R.S. Mendes, L.C. Malacarne, R.P.B. Santos, H.V. Ribeiro, S. Picoli, Earthquake-like patterns of acoustic emission in crumpled plastic sheets, EPL 92 (2010) 29001.

\bibitem{Ribeiro2} H.V. Ribeiro, L.S. Costa, L.G.A. Alves, P.A. Santoro, S. Picoli, E.K. Lenzi, R.S. Mendes, Analogies between the cracking noise of ethanol-dampened charcoal and earthquakes, Phys. Rev. Lett. 115 (2015) 025503.

\bibitem{Turner} D.L. Turner, Y. Shprits, M. Hartinger, V. Angelopoulos,  Explaining sudden losses of outer radiation belt electrons during geomagnetic storms, Nat. Phys. 8 (2012) 208.

\bibitem{Boettle} M. Boettle, D. Rybski, J.P. Kropp, How changing sea level extremes and protection measures alter coastal flood damages, Water Resour. Res. 49 (2013) 1199.

\bibitem{Rybski} D. Rybski, A. Holsten, J.P. Kropp, Towards a unified characterization of phenological phases: fluctuations and correlations with temperature, Physica A 390 (2011) 680.

\bibitem{Ribeiro} H.V. Ribeiro, F.J. Antonio, L.G.A. Alves, E.K. Lenzi, R.S. Mendes, Long-range spatial correlations and fluctuation statistics of lightning activity rates in Brazil, EPL 104 (2013) 69001.

\bibitem{Dove} M.R. Dove, D.M. Kammen, Science, society and the environment: applying anthropology and physics to sustainability, Routledge, New York, 2015.

\bibitem{WMO} World Meteorological Organization (WMO). Available: \url{https://www.wmo.int/pages/themes/climate/understanding_climate.php}. Accessed 2015 Jul.

\bibitem{Oliver} H.R. Oliver, S.A. Oliver (eds.), The role of water and the hydrological cycle in global change, Vol. 31 Springer, 1995.

\bibitem{Machiwal} D. Machiwal, M.K. Jha, Hydrologic time series analysis: theory and practice, Springer, New Delhi, 2012.

\bibitem{Hurst} H.E. Hurst, Long-term storage capacity of reservoirs, Transact. Am. Soc. Civil Eng. 116 (1951) 770.

\bibitem{Hosking} J.R.M. Hosking, Modeling persistence in hydrological time series using fractional differencing, Water Resour. Res. 20 (1984) 1898.

\bibitem{Hipel} K.W. Hipel, A.I. McLeod, Time series modeling of water resources and environmental systems. Developments in water science. Elsevier Science, Amsterdam, 1994.

\bibitem{Montanari2} A. Montanari, R. Rosso, M.S. Taqqu, Fractionally differenced ARIMA models applied to hydrologic time series: Identification, estimation, and simulation. Water Resour. Res. 33 (1997) 1035.

\bibitem{Koutsoyiannis} D. Koutsoyiannis, The Hurst phenomenon and fractional Gaussian noise made easy, Hydrolog. Sci. J. 47 (2002) 573.

\bibitem{Dahlstedt} K. Dahlstedt, H.J. Jensen, Fluctuation spectrum and size scaling of river flow and level, Physica A 348 (2005) 596.

\bibitem{Wang} W. Wang,  P. H. A. J. M. Van Gelder, J. K.  Vrijling, X. Chen, Detecting long-memory: Monte Carlo simulations and application to daily stream flow processes. Hydrol. Earth Syst. Sci. 11 (2007) 851.

\bibitem{Dolgonosov} B.M. Dolgonosov, K.A. Korchagin, N.V. Kirpichnikova, Modeling of annual oscillations and 1/f noise of daily river discharges, J. Hydrol. 357 (2008) 174.

\bibitem{Movaheda} M.S. Movahed, E. Hermanis, Fractal analysis of river flow fluctuations, Physica A 387 (2008) 915.

\bibitem{Zhang} Q. Zhang, C.-Y. Xu, Z. Yu, C.-L. Liu, Y.D. Chen, Physica A 388 (2009) 927.

\bibitem{Zhang2} Q. Zhang, C.-Y. Xu, T. Yang, Multifractal analyses of daily rainfall time series in Pearl River basin of China, Stoch. Environ. Res. Risk. Assess. 23 (2009) 1103.

\bibitem{Montanari3} A. Montanari, R. Rosso, M.S Taqqu, A seasonal fractional ARIMA model applied to the Nile River monthly flows at Aswan. Water Resour. Res. 36 (2010) 1249.

\bibitem{Domenico} M. De Domenico, V. Latora, Scaling and universality in river flow dynamics, EPL 94 (2011) 58002.

\bibitem{Yu} Z.-G. Yu, Y. Leung, Y.D. Chen, Q. Zhang, V. Anh, Y. Zhou, Multifractal analyses of daily rainfall time series in Pearl River basin of China, Physica A 405 (2014) 193.

\bibitem{Montanari} A. Montanari, Hydrology of the Po River: looking for changing patterns in river discharge, Hydrol. Earth Syst. Sci. 16 (2012) 741, 3739.

\bibitem{Tessier} Y. Tessier, S. Lovejoy, P. Hubert, D. Schertzer, S. Pecknold, Multifractal analysis and modeling of rainfall and river flows and scaling, causal transfer functions, J. Geophys. Res. 101 (1996) 26427.

\bibitem{Kantelhardt2} J.W. Kantelhardt, D. Rybski, S.A. Zschiegner, P. Braun, E. Koscielny-Bunde, V. Livina, S. Havlin, A. Bunde, Multifractality of river runoff and precipitation: comparison of fluctuation analysis and wavelet methods, Physica A 330 (2003) 240.

\bibitem{Bogachev} M.I. Bogachev, A. Bunde, Universality in the precipitation and river runoff, EPL 97 (2012) 48011.

\bibitem{Hajian} S. Hajian, M.S. Movahed, Multifractal detrended cross-correlation analysis of sunspot numbers and river flow fluctuations, Physica A 389 (2010) 4942.

\bibitem{Janosi} I.M. J\'anosi, J.A.C. Gallas, Growth of companies and water-level fluctuations of the river Danube, Physica A 271 (1999) 448.

\bibitem{Bramwell} S.T. Bramwell, T. Fennell, P.C.W. Holdsworth, B. Portelli, Universal Fluctuations of the Danube Water Level: a Link with Turbulence, Criticality and Company Growth, EPL 57 (2002) 310.

\bibitem{Porporato} A. Porporato, L. Ridolfi, Nonlinear analysis of river flow time sequences, Water Resour. Res. 33 (1997) 1353.

\bibitem{Bordignon} S. Bordignon, F. Lisi, Nonlinear analysis and prediction of river flow time series, Environmetrics 11 (2000) 463.

\bibitem{Livina} V. Livina, Y. Ashkenazy, Z. Kizner, V. Strygin, A. Bunde, S. Havlin, A stochastic model of river discharge fluctuations, Physica A 330 (2003) 283.

\bibitem{Mihailovic} D.T. Mihailovi\'c, E. Nikoli\'c-Dori\'c, N. Dreskovi\'c, G. Mimi\'c, Complexity analysis of the turbulent environmental fluid flow time series, Physica A 395 (2014) 96.

\bibitem{Hauhs} M. Hauhs, H. Lange, Classification of runoff in headwater catchments: a physical problem, Geogr. Compass 2 (2008) 235.

\bibitem{Zunino2} L. Zunino, M.C. Soriano, O.A. Rosso, Distinguishing chaotic and stochastic dynamics from time series by using a multiscale symbolic approach, Phys. Rev. E 86 (2012) 046210.

\bibitem{Lange} H. Lange, O.A. Rosso, M. Hauhs, Ordinal pattern and statistical complexity analysis of daily stream flow time series, Eur. Phys. J. Spec. Top. 222 (2013) 535.

\bibitem{Serinaldi} F. Serinaldi, L. Zunino, O. Rosso, Complexity-entropy analysis of daily stream flow time series in the continental United States, Stoch. Env. Res. Risk A. 28 (2014) 1685.

\bibitem{Jha} S.K. Jha, H. Zhao, F.M. Woldemeskel, B. Sivakumar, Network theory and spatial rainfall connections: An interpretation, Journal of Hydrology 527 (2015) 13.

\bibitem{Scarsoglio} S. Scarsoglio, F. Laio, L. Ridolfi, Climate Dynamics: A Network-Based Approach for the Analysis of Global Precipitation, PLoS One 8 (2013) e71129.

\bibitem{Sivakumar1} B. Sivakumar, Networks: a generic theory for hydrology?, Stoch. Environ. Res. Risk. Assess. 29 (2015) 761.

\bibitem{Sivakumar2} B. Sivakumar, F.M. Woldemeskel, Complex networks for streamflow dynamics, Hydrol. Earth Syst. Sci. Discuss. 11 (2014) 7255.

\bibitem{Sivakumar3} B. Sivakumar, F.M. Woldemeskel,  A network-based analysis of spatial rainfall connections, Environ. Modell. Soft. 69 (2015) 55.

\bibitem{Luque} B. Luque, L. Lacasa, F. Ballesteros, J. Luque, Horizontal visibility graphs: Exact results for random time series, Phys. Rev. E 80 (2009) 046103.

\bibitem{Lacasa} L. Lacasa, B. Luque, F. Ballesteros, J. Luque, J. Nu\~no. From time series to complex networks: The visibility graph, Proc. Natl. Acad. Sci. USA 105 (2008) 4972.

\bibitem{Lacasa2} L. Lacasa, R. Toral, Description of stochastic and chaotic series using visibility graphs, Phys. Rev. E 82 (2010) 036120.

\bibitem{ONS} Operador Nacional do Sistema El\'etrico (ONS). Available: \url{http://www.ons.org.br/operacao/vazoes_naturais.aspx}. Accessed 2015 Jan.

\bibitem{Lacasa3} L. Lacasa, B. Luque, J. Luque, N.C. Nuno, The visibility graph: A new method for estimating the Hurst exponent of fractional Brownian motion, EPL 86 (2009) 30001.

\bibitem{Yang} Y. Yang, J.B. Wang, H.J. Yang, J.S. Mang, Visibility graph approach to exchange rate series, Physica A. 388 (2009) 4431.

\bibitem{Elsner} J.B. Elsner, T.H. Jagger, E.A. Fogarty, Visibility network of United States hurricanes, Geophys. Res. Lett. 36 (2009) L16702.

\bibitem{Ahmadlou} M. Ahmadlou, H. Adeli, A. Adeli, New diagnostic EEG markers of the Alzheimer's disease using visibility graph, J. Neural Transm. 117 (2010) 1099.

\bibitem{Murks} A. Murks, M. Perc, Evolutionary games on visibility graphs, Adv. Complex Syst. 14 (2011) 307. 

\bibitem{Telesca} L. Telesca, M. Lovallo, Analysis of seismic sequences by using the method of visibility graph, EPL 97 (2012) 50002.

\bibitem{Gao} Z.K. Gao, N.D. Jin, Characterization of chaotic dynamic behavior in the gas-liquid slug flow using directed weighted complex network analysis, Physica A. 391 (2012) 3005.

\bibitem{Jiang} S. Jiang, C.H. Bian, X.B. Ning, Q.L.D.Y. Ma, Visibility graph analysis on heartbeat dynamics of meditation training, Appl. Phys. Lett. 102 (2013) 253702.

\bibitem{Zhuang} E. Zhuang, M. Small, G. Feng, Time series analysis of the developed financial markets' integration using visibility graphs, Physica A. 410 (2014) 483.

\bibitem{Telesca2} L. Telesca, M. Lovallo, L. Toth, Visibility graph analysis of 2002-2011 Pannonian seismicity, Physica A. 416 (2014) 219.

\bibitem{Zou} Y. Zou, R.V. Donner, N. Marwan, M. Small, J. Kurths, Long-term changes in the north-south asymmetry of solar activity: a nonlinear dynamics characterization using visibility graphs,  Nonlin. Processes Geophys. 21 (2014) 1113.

\bibitem{Zhangvg} B. Zhang, J. Wang, W. Fang, Volatility behavior of visibility graph EMD financial time series from Ising interacting system, Physica A. 432 (2015) 301.

\bibitem{Koutsoyiannis2} D. Koutsoyiannis, Nonstationarity versus scaling in hydrology, Journal of Hydrology 324 (2006) 239.

\bibitem{Efron} B. Efron, R. Tibshirani, An Introduction to the Bootstrap, Chapman \& Hall, New York, 1993.




\bibitem{Watts} D.J. Watts, S.H. Strogatz, Collective dynamics of `small-world' networks, Nature 393 (1998) 440.

\bibitem{Newman} M.E.J. Newman, S.H. Strogatz, D.J. Watts, Random graphs with arbitrary degree distributions and their applications, Phys. Rev. E 64 (2001) 026118.

\bibitem{Newman2} M.E.J. Newman, The structure and function of complex networks, SIAM Review 45 (2003) 167256.

\end{thebibliography}
\end{document}